\documentstyle[times,psfig,mathrsfs]{mn}

\newif\ifAMStwofonts
\AMStwofontstrue

\ifoldfss
  \newcommand{\rmn}[1] {{\rm #1}}

  \ifCUPmtlplainloaded \else
    \NewTextAlphabet{textbfit} {cmbxti10} {}
    \NewTextAlphabet{textbfss} {cmssbx10} {}
    \NewMathAlphabet{mathbfit} {cmbxti10} {} % for math mode
    \NewMathAlphabet{mathbfss} {cmssbx10} {} %  "   "    "
  \fi
  \ifAMStwofonts
    \ifCUPmtlplainloaded \else
      \NewSymbolFont{upmath} {eurm10}
      \NewSymbolFont{AMSa} {msam10}
      \NewMathSymbol{\upi}     {0}{upmath}{19}
      \NewMathSymbol{\umu}     {0}{upmath}{16}
      \NewMathSymbol{\upartial}{0}{upmath}{40}
      \NewMathSymbol{\leqslant}{3}{AMSa}{36}
      \NewMathSymbol{\geqslant}{3}{AMSa}{3E}

    \fi
  \fi
\fi % End of OFSS

\ifnfssone
  \newmathalphabet{\mathit}
  \addtoversion{normal}{\mathit}{cmr}{m}{it}
  \addtoversion{bold}{\mathit}{cmr}{bx}{it}
  \newcommand{\rmn}[1] {\mathrm{#1}}

  \newmathalphabet{\mathbfit} % math mode version of \textbfit{..}
  \addtoversion{normal}{\mathbfit}{cmr}{bx}{it}
  \addtoversion{bold}{\mathbfit}{cmr}{bx}{it}
  \newmathalphabet{\mathbfss} % math mode version of \textbfss{..}
  \addtoversion{normal}{\mathbfss}{cmss}{bx}{n}
  \addtoversion{bold}{\mathbfss}{cmss}{bx}{n}
  \ifAMStwofonts
    \ifCUPmtlplainloaded \else
      \UseAMStwoboldmath
      \makeatletter
      \new@mathgroup\upmath@group
      \define@mathgroup\mv@normal\upmath@group{eur}{m}{n}
      \define@mathgroup\mv@bold\upmath@group{eur}{b}{n}
      \edef\UPM{\hexnumber\upmath@group}
      \new@mathgroup\amsa@group
      \define@mathgroup\mv@normal\amsa@group{msa}{m}{n}
      \define@mathgroup\mv@bold\amsa@group{msa}{m}{n}
      \edef\AMSa{\hexnumber\amsa@group}
      \makeatother
      \mathchardef\upi="0\UPM19
      \mathchardef\umu="0\UPM16
      \mathchardef\upartial="0\UPM40
      \mathchardef\leqslant="3\AMSa36
      \mathchardef\geqslant="3\AMSa3E
    \fi
  \fi
\fi % End of NFSS release 1

\ifnfsstwo
  \newcommand{\rmn}[1] {\mathrm{#1}}

  \DeclareMathAlphabet{\mathbfit}{OT1}{cmr}{bx}{it}
  \SetMathAlphabet\mathbfit{bold}{OT1}{cmr}{bx}{it}
  \DeclareMathAlphabet{\mathbfss}{OT1}{cmss}{bx}{n}
  \SetMathAlphabet\mathbfss{bold}{OT1}{cmss}{bx}{n}
  \ifAMStwofonts
    \ifCUPmtlplainloaded \else
      \DeclareSymbolFont{UPM}{U}{eur}{m}{n}
      \SetSymbolFont{UPM}{bold}{U}{eur}{b}{n}
      \DeclareSymbolFont{AMSa}{U}{msa}{m}{n}
      \DeclareMathSymbol{\upi}{0}{UPM}{"19}
      \DeclareMathSymbol{\umu}{0}{UPM}{"16}
      \DeclareMathSymbol{\upartial}{0}{UPM}{"40}
      \DeclareMathSymbol{\leqslant}{3}{AMSa}{"36}
      \DeclareMathSymbol{\geqslant}{3}{AMSa}{"3E}
    \fi
  \fi
\fi % End of NFSS release 2

\ifCUPmtlplainloaded \else
  \ifAMStwofonts \else % If no AMS fonts
    \def\upi{\pi}
    \def\umu{\mu}
    \def\upartial{\partial}
  \fi
\fi

   \title[Manganese evolution in Omega Cen]{Manganese evolution in Omega 
     Centauri: a clue to the cluster formation mechanisms?}
   \author[D. Romano et al.]{Donatella Romano,$^{1, 2}$\thanks{E-mail: 
           donatella.romano@oabo.inaf.it} Gabriele Cescutti$^{3, 4}$ and 
         Francesca Matteucci$^{4, 5}$\\
         $^{1}$INAF, Osservatorio Astronomico di Bologna,
               Via Ranzani 1, I-40127 Bologna, Italy\\
         $^{2}$Dipartimento di Astronomia, Universit\`a di Bologna,
               Via Ranzani 1, I-40127 Bologna, Italy\\
         $^{3}$Laboratoire d'Astrophysique, \'Ecole Polytechnique F\'ed\'erale 
               de Lausanne (EPFL), Observatoire, 1290 Sauverny, Switzerland\\
         $^{4}$Dipartimento di Fisica, Universit\`a di Trieste,
               Via Tiepolo 11, I-34143 Trieste, Italy\\
         $^{5}$INAF, Osservatorio Astronomico di Trieste,
               Via Tiepolo 11, I-34143 Trieste, Italy}
   
\begin{document}

     \date{Accepted 2011 July 27. Received 2011 July 26; in original form 2011 
       March 30}

     \pagerange{\pageref{firstpage}--\pageref{lastpage}} \pubyear{2011}

     \maketitle

     \label{firstpage}

%%%%%%%%%%%%%%%%%%%%%%%%%%%%%%%%%%%%%%%%%%%%%%%%

   \begin{abstract}
     We model the evolution of manganese relative to iron in the progenitor 
     system of the globular cluster Omega Centauri by means of a 
     self-consistent chemical evolution model. We use stellar yields that 
     already reproduce the measurements of [Mn/Fe] versus [Fe/H] in Galactic 
     field disc and halo stars, in Galactic bulge stars and in the Sagittarius 
     dwarf spheroidal galaxy. We compare our model predictions to the Mn 
     abundances measured in a sample of 10 red giant members and 6 subgiant 
     members of $\omega$\,Cen. The low values of [Mn/Fe] observed in a few, 
     metal-rich stars of the sample cannot be explained in the framework of our 
     standard, homogeneous chemical evolution model. Introducing cooling flows 
     that selectively bring to the cluster core only the ejecta from specific 
     categories of stars does not help to heal the disagreement with the 
     observations. The capture of field stars does not offer a viable 
     explanation either. The observed spread in the data and the lowest [Mn/Fe] 
     values could, in principle, be understood if the system experienced 
     inhomogeneous chemical evolution. Such an eventuality is qualitatively 
     discussed in the present paper. However, more measurements of Mn in 
     $\omega$\,Cen stars are needed to settle the issue of Mn evolution in this 
     cluster.
    \end{abstract}


%%%%%%%%%%%%%%%%%%%%%%%%%%%%%%%%%%%%%%%%%%%%%%%%

   \section{Introduction}
   \label{sec:int}

   The abundance ratios of chemical elements that are produced on different 
   time scales by stars of different masses are essential probes of the history 
   of chemical enrichment in different types of stellar populations. They tell 
   which stars -- and in which proportions -- have contributed to the chemical 
   evolution of a given system at any time. After Tinsley (1979), it has 
   become customary to explain the trend of oxygen relative to iron as due to 
   the different roles played by type Ia supernovae (SNeIa) and type II 
   supernovae (SNeII) in the chemical enrichment of galaxies (see also Greggio 
   \& Renzini 1983; Matteucci \& Greggio 1986). This interpretation is known as 
   the `time-delay model'. Oxygen is produced mostly by core-collapse SNe on 
   very short time scales, of the order of a few Myr to tenths of Myr. The bulk 
   of iron, instead, comes later from SNIa explosions, on time scales ranging 
   from 30~Myr to a Hubble time. Hence, the [O/Fe] ratio in a given system is 
   high as long as SNeII dominate the interstellar medium (ISM) pollution, then 
   decreases to solar and subsolar values when SNeIa become the major 
   contributors to the production of Fe. The maximum SNIa rate depends on the 
   adopted progenitor model, as well as on the assumed star formation history 
   (Matteucci \& Recchi 2001); thus, the measurements of [O/Fe] are a powerful 
   diagnostics of the star formation history in galaxies.

   Of particular interest are those elements whose yields depend on the 
   metallicity of the parent stars; manganese (Mn) is one of them. McWilliam, 
   Rich \& Smecker-Hane (2003) confronted the [Mn/Fe] versus [Fe/H] relation in 
   the Galactic bulge, in the solar neighbourhood and in the Sagittarius dwarf 
   spheroidal galaxy (Sgr dSph). They suggested that the Mn yields from both 
   SNeIa and SNeII are metallicity dependent. Their proposals were in agreement 
   with extant nucleosynthesis calculations for massive stars (e.g. Arnett 
   1971; Woosley \& Weaver 1995), but there was not modelling of explosive 
   nucleosynthesis in SNeIa supporting their findings.

   Shetrone et al. (2003) measured the Mn abundances of a dozen individual red 
   giant stars in the Sculptor, Fornax, Carina and Leo~I dSphs. They also 
   conclude that SNIa contributions to the synthesis of Mn must be metallicity 
   dependent, with very little Mn produced until [Fe/H]~= $-$1.

   Later on, Ohkubo et al. (2006) showed that, indeed, [Mn/Fe] -- and [Ni/Fe] 
   -- in the ejecta of SNeIa depend on metallicity of SNIa progenitors. 
   Furthermore, Cescutti et al. (2008) demonstrated, by means of 
   self-consistent chemical evolution models, that the run of [Mn/Fe] with 
   [Fe/H] in the three independent stellar systems -- the Galactic bulge, the 
   solar neighbourhood and the Sgr dSph-- can be understood \emph{only in 
     terms of a metallicity-dependent yield of Mn} from SNeIa. The time-delay 
   model alone is insufficient to explain the behaviour of [Mn/Fe] versus 
   [Fe/H] in the three systems. Cescutti et al. (2008) propose that the yield 
   of Mn in SNeIa increases with the metallicity of the progenitors, 
   $Y_{\rmn Mn} (Z) \propto Z^{0.65}$. This is in agreement with the tight 
   correlation between the Mn-to-Cr mass ratio in the ejecta of SNeIa and the 
   metallicity of the progenitor found by Badenes, Bravo \& Hughes (2008), 
   $M_{\rmn Mn}/M_{\rmn Cr}$~= 5.3~$\times$~$Z^{0.65}$.

   More recently, Mn abundances have been studied for the first time on a 
   significant metallicity range in the peculiar globular cluster 
   $\omega$\,Centauri (Cunha et al. 2010; Pancino et al. 2011). In the 
   metal-poor regime, Cunha et al. (2010) find that the LTE values of [Mn/Fe] 
   in $\omega$\,Cen stars overlap those of their solar neighbourhood analogues. 
   However, at variance with the solar neighbourhood trend, [Mn/Fe] declines in 
   more metal-rich stars (see also Pancino et al. 2011). Non-LTE calculations 
   confirm the conclusion of a well distinct pattern of [Mn/Fe] versus [Fe/H] 
   in $\omega$\,Cen (Cunha et al. 2010). It is worth noticing that, in the 
   metallicity range $-$2.7~$<$~[Fe/H] $< -$0.7, all other Galactic globulars 
   have Mn abundances equivalent to those of halo field stars (Sobeck et al. 
   2006).

   Although historically classified as a globular cluster, $\omega$\,Cen is 
   possibly the naked nucleus of a small galactic satellite captured by the 
   Milky Way many Gyr ago (e.g. Freeman 1993; Bekki \& Freeman 2003). As such, 
   it likely suffered a complex chemical enrichment history, marked by the 
   occurrence of strong differential galactic winds (Romano et al. 2007, 
   2010a). The challenging bet is: can the low values of [Mn/Fe] observed in 
   the more metal-rich stars of $\omega$\,Cen be explained in the context of 
   its peculiar evolutive history?

   In this paper, we contrast the evolution of Mn in the Milky Way with that in 
   $\omega$\,Cen. The goal of this work is twofold. First, we want to add the 
   analysis of another distinct stellar population to previous theoretical 
   study of the evolution of Mn in different environments by Cescutti et al. 
   (2008). Second, we want to get better insight into the mechanisms of 
   formation and evolution of $\omega$\,Cen. The predictions of our models for 
   the Milky Way and for $\omega$\,Cen are compared to both LTE and non-LTE Mn 
   abundances in the two systems. The paper is organized as follows: in 
   Sect.~2, we briefly review the relevant observational data. In Sect.~3, we 
   describe the adopted chemical evolution models. In Sect.~4, we discuss the 
   model results. In Sect.~5, we draw our conclusions.

   \section{Observational data}

\begin{figure*}
     \psfig{figure=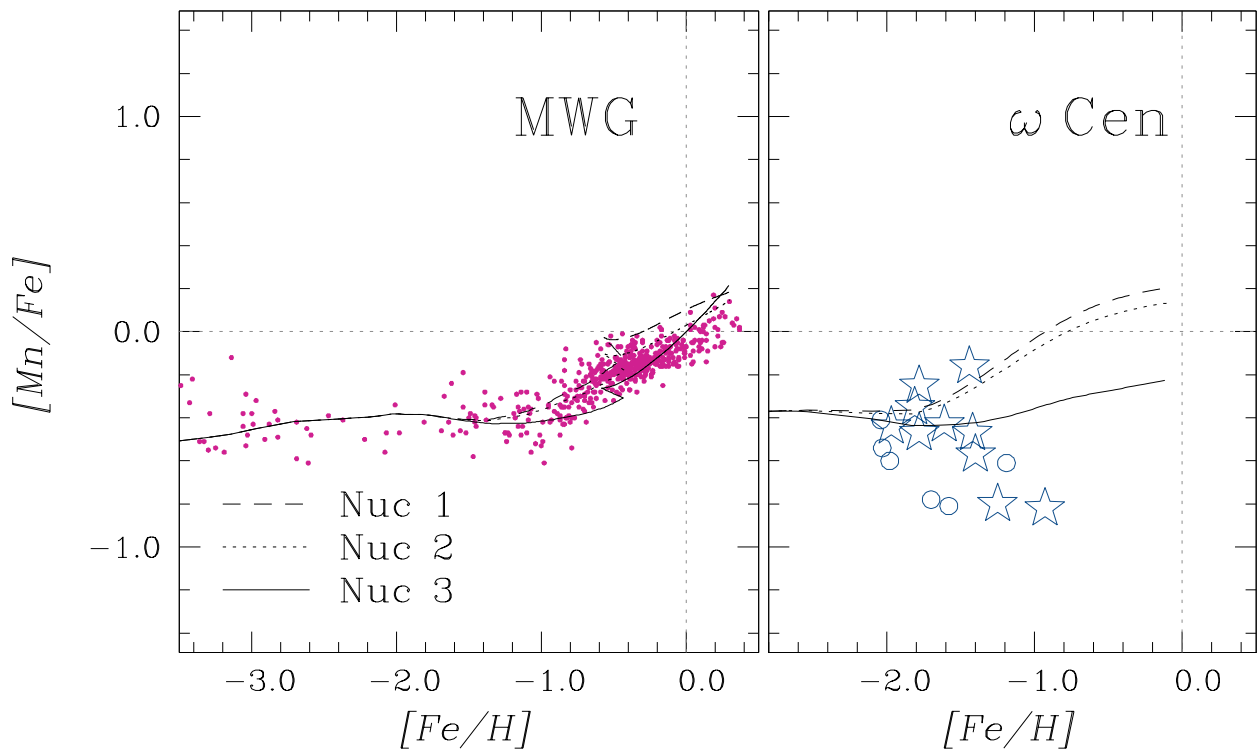,width=\textwidth}
     \caption{ [Mn/Fe] versus [Fe/H] relation in the solar neighbourhood (left 
       panel) and in $\omega$\,Cen (right panel). The dashed curves in both 
       panels are the theoretical trends obtained with the 
       metallicity-dependent yields of Woosley \& Weaver (1995) for SNeII and 
       the yields of Iwamoto et al. (1999) for solar-metallicity SNeIa at all 
       metallicities. The dotted and solid curves show the effect of taking 
       into account in two different ways (see text) the metallicity dependence 
       of the Mn yield from SNeIa. Left panel: filled circles are LTE data from 
       several sources (see Sect.~2 for references). Right panel: LTE data are 
       taken from Cunha et al. (2010, stars) and Pancino et al. (2011, open 
       circles).
     }
     \label{fig:lte}
   \end{figure*}

   For the purpose of comparison with the results of our models, we use LTE 
   Mn\,{\sc I} abundances of Galactic stars by Cayrel et al. (2004), Gratton et 
   al. (2003), Reddy et al. (2003), Reddy, Lambert \& Allende Prieto (2006) and 
   Feltzing, Fohlman \& Bensby (2007). These studies are selected to cover the 
   whole metallicity range of solar neighbourhood stars. Other recent 
   measurements of Mn in Galactic disc and halo stars can be found in Nissen et 
   al. (2000), Prochaska \& McWilliam (2000), Norris, Ryan \& Beers (2001), 
   Carretta et al. (2002), Bai et al. (2004) and Lai et al. (2008). For 
   $\omega$\,Cen, we use LTE data from Cunha et al. (2010, 10 stars) and 
   Pancino et al. (2011, 6 stars). While Cunha et al. (2010) analysed Mn lines 
   around 6000~\AA, Pancino et al. (2011) analysed Mn lines around 4000~\AA. 
   This could explain the marginal disagreement between the two data sets and 
   lead to an artificial increase of the scatter in the data when the two data 
   sets are plotted together (Fig.~\ref{fig:lte}, right panel). Cohen (1981) 
   and Gratton (1982) also published Mn abundances of $\omega$\,Cen stars. 
   However, their determinations do not take hyperfine splitting of spectral 
   lines into account and are, thus, not considered here.

   Despite the adoption of different methods and assumptions to determine the 
   Mn abundances, all the studies mentioned above conclude that Mn is deficient 
   compared to iron ([Mn/Fe]~$\simeq -$0.5) in metal-poor stars and that 
   [Mn/Fe] increases with increasing [Fe/H] from the halo to the thin disc 
   populations. The stars in $\omega$\,Cen present a trend at odds with that of 
   their solar neighbourhood counterparts, i.e. a [Mn/Fe] ratio declining with 
   increasing [Fe/H] (Cunha et al. 2010). Notice, however, that this statement 
   rests with the determination of Mn in only two stars.

   In the LTE approximation, the abundances based on Mn\,{\sc I} lines could be 
   underestimated by as much as 0.4 dex at low metallicities, with non-LTE 
   effects being less pronounced in high-metallicity stars (Bergemann \& Gehren 
   2008). Up to now, non-LTE Mn abundances have been computed only for a few 
   stars in the Milky Way (Bergemann \& Gehren 2008) and in $\omega$\,Cen 
   (Cunha et al. 2010). When the non-LTE corrections are applied to solar 
   neighbourhood stars, a shallower rise is found from slightly subsolar values 
   in the halo ([Mn/Fe]~$\simeq -$0.1) to solar ratios in the thin disc. The 
   non-LTE abundance analysis of $\omega$\,Cen stars fully confirms the odd 
   behaviour of decreasing Mn with increasing [Fe/H] depicted by LTE studies 
   (see discussion in Cunha et al. 2010), but once again the result bases on 
   the analysis of only a handful of stars. We discuss further the non-LTE Mn 
   abundances in comparison to our model results in Sect.~4.

   \section{The chemical evolution models}

   The chemical evolution model for the solar neighbourhood is basically the 
   `two-infall model' of Chiappini, Matteucci \& Gratton (1997), except for the 
   adopted stellar lifetimes, nucleosynthesis and initial mass function (IMF; 
   see Romano et al. 2005, 2010b). In particular, the Kroupa, Tout \& Gilmore 
   (1993) IMF is assumed instead of the Scalo (1986) one. This makes the 
   predicted present-day SNIa-to-SNII rate ratio to agree with the observations 
   (Romano et al. 2005, their table~4) and allows to better explain the 
   evolution of deuterium in the Galaxy (Romano 2010).

   The adopted model for $\omega$\,Cen assumes that this cluster was once 
   located at the centre of a more massive system, that evolved in isolation 
   before being accreted and almost totally disrupted by the interaction with 
   the Milky Way. Because of the shallow potential well, the chemical evolution 
   of the original system -- and of the embedded proto-cluster -- turns out to 
   be significantly affected by galactic outflows triggered by multiple SN 
   explosions (Romano et al. 2007). These outflows deprive the original system 
   of a large fraction of its metals, thus allowing the observed Na-O 
   anticorrelation and the extreme level of He enhancement to set up in a 
   (minor) fraction of the hosted stars (Romano et al. 2010a).

   For details about the basic assumptions and equations of the models, we 
   refer the reader to the papers quoted above. As for the adopted 
   nucleosynthesis prescriptions, they can be found below.

%%%%%%%%%%%%%%%%%%%%%%%%%%%%%%%%%%%%%%%%%%%%%%%% TWO COLUMN FIGURE
%
%
%%%%%%%%%%%%%%%%%%%%%%%%%%%%%%%%%%%%%%%%%%%%%%%%

%%%%%%%%%%%%%%%%%%%%%%%%%%%%%%%%%%%%%%%%%%%%%%%% TWO COLUMN FIGURE
%
   \begin{figure*}
     \psfig{figure=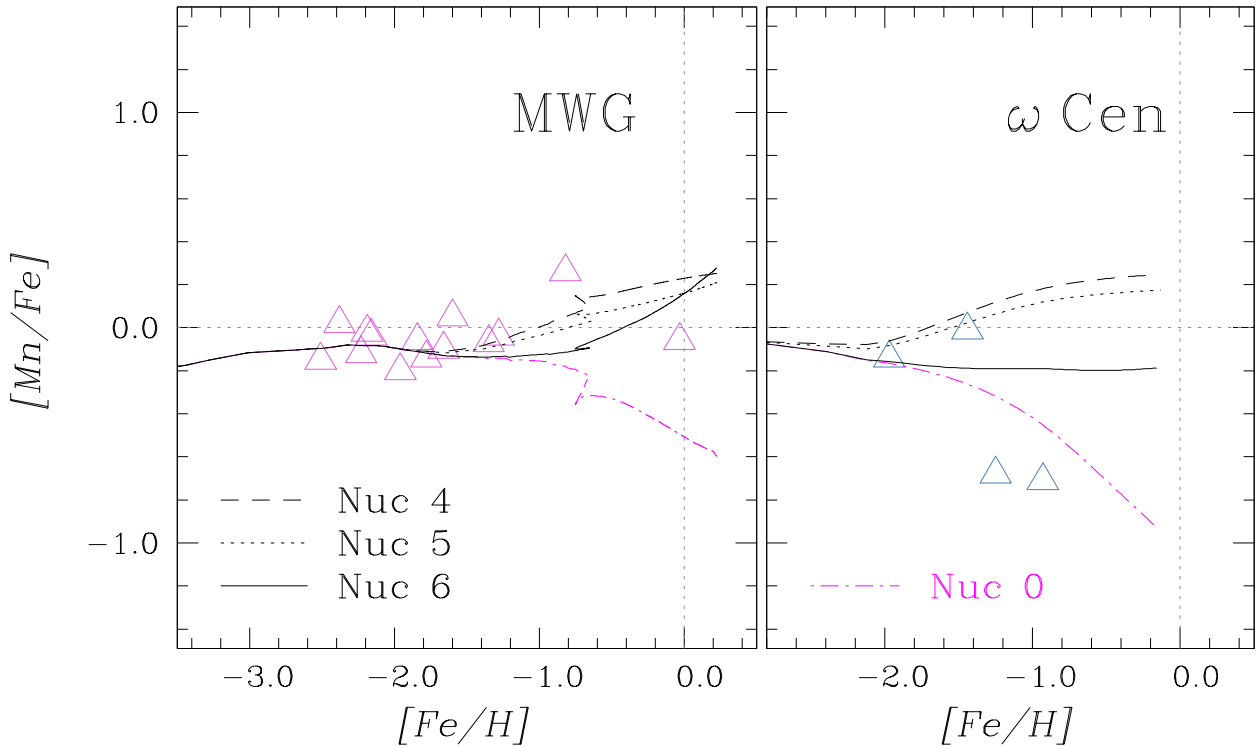,width=\textwidth}
     \caption{ [Mn/Fe] versus [Fe/H] relation in the solar neighbourhood (left 
       panel) and in $\omega$\,Cen (right panel). The dashed curves in both 
       panels refer to the predictions of models adopting the 
       metallicity-dependent yields of Woosley \& Weaver (1995) with Fe yields 
       halved for SNeII and the yields of Iwamoto et al. (1999) for 
       solar-metallicity SNeIa at all metallicities. The dotted and solid 
       curves show the effect of taking into account the metallicity dependence 
       of the Mn yield from SNeIa. The dot-dashed curves show the predictions 
       of models computed with zero Mn production from SNeIa. Open triangles 
       are non-LTE data (references are given in Sect.~2).
     }
     \label{fig:nlte}
   \end{figure*}
%
%%%%%%%%%%%%%%%%%%%%%%%%%%%%%%%%%%%%%%%%%%%%%%%%

   \subsection{Nucleosynthesis prescriptions}

   As already mentioned, Fe and Mn are produced by both SNeII and SNeIa, though 
   in different proportions. In our computations, we explore the consequences 
   of adopting six different prescriptions about the synthesis of Fe and Mn in 
   stars:
   \begin{enumerate}
     \item Nuc~1: we use the metal-dependent yields of Woosley \& Weaver (1995) 
       for SNeII and the yields of Iwamoto et al. (1999) for solar-metallicity 
       SNeIa (their model W7) at all metallicities.
     \item Nuc~2: as above, but we interpolate between zero-metallicity and 
       solar-metallicity SNIa yields (models W70 and W7 in Iwamoto et al. 1999).
     \item Nuc~3: as above, but we modify the Mn yield from SNeIa following 
       Cescutti et al. (2008):
       \begin{equation}
         Y_{\rmn Mn}(Z) = Y_{\rmn Mn}^{\rmn W7}\bigg ( 
         \frac{Z}{Z_{\sun}}\bigg )^{0.65},
         \label{eq:MnZ}
       \end{equation}
       where $Z$ and $Z_{\sun}$ are the metallicities of the SNIa 
       progenitor and of the Sun at birth, respectively.
     \item Nuc~4: we use the metal-dependent yields of Woosley \& Weaver (1995) 
       \emph{with Fe yields halved} for SNeII and the yields of Iwamoto et al. 
       (1999) for solar-metallicity SNeIa (their model W7) at all metallicities.
     \item Nuc~5: as above, but we interpolate between zero-metallicity and 
       solar-metallicity SNIa yields (models W70 and W7 in Iwamoto et al. 1999).
     \item Nuc~6: as above, but with a metal-dependent yield of Mn from SNeIa 
       as in Eq.~\ref{eq:MnZ}.
   \end{enumerate}
   Models without Mn production from SNeIa were computed as well (models 
   labeled Nuc~0) and are discussed in Sect.~4.2.

   The first and third choice are in common with Cescutti et al. (2008), who 
   conclude that the Mn yield from SNeIa must be metallicity-dependent. Their 
   result derives from the simultaneous analysis of the behaviour of [Mn/Fe] 
   versus [Fe/H] in the Galactic bulge, in the solar neighbourhood and in the 
   Sgr dSph. Here, we add the study of Mn evolution in $\omega$\,Cen to their 
   survey.

   Notice that Cescutti et al. (2008) adopted model W7 of Iwamoto et al. (1999) 
   at all metallicities. This was done because the physics -- and 
   nucleosynthesis output especially for those elements produced in the inner 
   part of the WD -- of metal-free SNeIa are likely to be not significantly 
   different from those at $Z \ne$ 0. Also, only a tiny fraction of all SNeIa 
   that explode in a galaxy forms from truly $Z$~= 0 matter. In fact, a few 
   SNII explosions suffice to rise the metal content of the ISM from zero to 
   non-zero. On the other hand, the adoption of an empirical law of the form 
   suggested by Cescutti et al. (2008) for Mn was justified by the fact that it 
   could reproduce the Mn evolution in different objects and that Mn is one of 
   those elements produced in the external layers of the WD and therefore more 
   dependent on the initial metallicity of the progenitor of the C-O WD 
   (Thielemann, private communication).

   At this point, it is worth mentioning once again that Badenes et al. (2008) 
   also find that the Mn yield from SNeIa declines with decreasing metallicity. 
   The metallicity dependence they suggest is the same as in Eq.~\ref{eq:MnZ}. 
   Their result, which springs from detailed SNIa modelling, is stable against 
   variations in the initial conditions and explosion mechanisms (either 
   delayed detonation or deflagration) of the models explored.

   \section{Model results and discussion}

%%%%%%%%%%%%%%%%%%%%%%%%%%%%%%%%%%%%%%%%%%%%%%%% TWO COLUMN FIGURE
%
   \begin{figure*}
     \psfig{figure=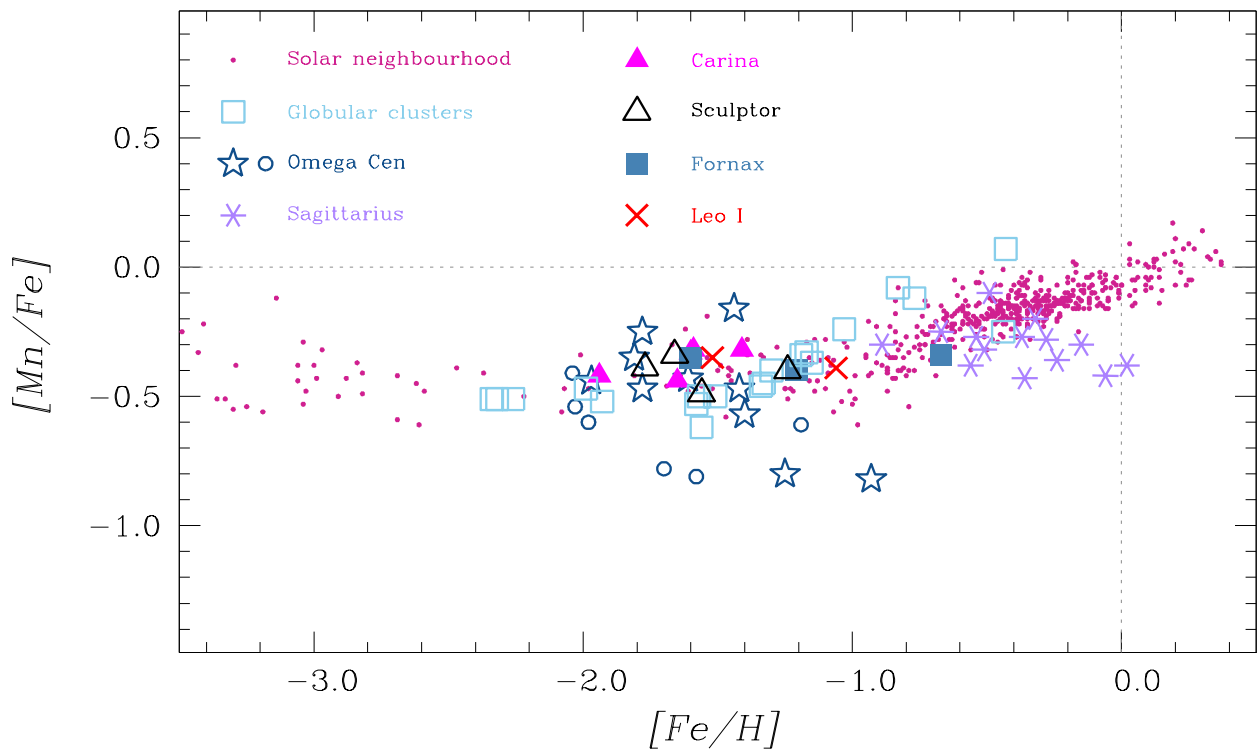,width=\textwidth}
     \caption{ Observational [Mn/Fe] versus [Fe/H] relationships for field 
       stars in the solar neighbourhood (filled circles; Cayrel et al. 2004; 
       Gratton et al. 2003; Reddy et al. 2003, 2006; Feltzing et al. 2007), 20 
       Galactic globular clusters (open squares; Gratton et al. 2006; Carretta 
       et al. 2007; Carretta 2010, private communication), $\omega$\,Cen 
       (stars: Cunha et al. 2010; open circles: Pancino et al. 2011), 
       Sagittarius (main body and Terzan~7, asterisks; Sbordone et al. 2007) 
       and 4 dSphs of the Local Group (filled triangles: Carina; open 
       triangles: Sculptor; filled squares: Fornax; crosses: Leo~I; Shetrone et 
       al. 2003).
     }
     \label{fig:all}
   \end{figure*}
%
%%%%%%%%%%%%%%%%%%%%%%%%%%%%%%%%%%%%%%%%%%%%%%%%

%%%%%%%%%%%%%%%%%%%%%%%%%%%%%%%%%%%%%%%%%%%%%%%% TWO COLUMN FIGURE
%
   \begin{figure*}
     \psfig{figure=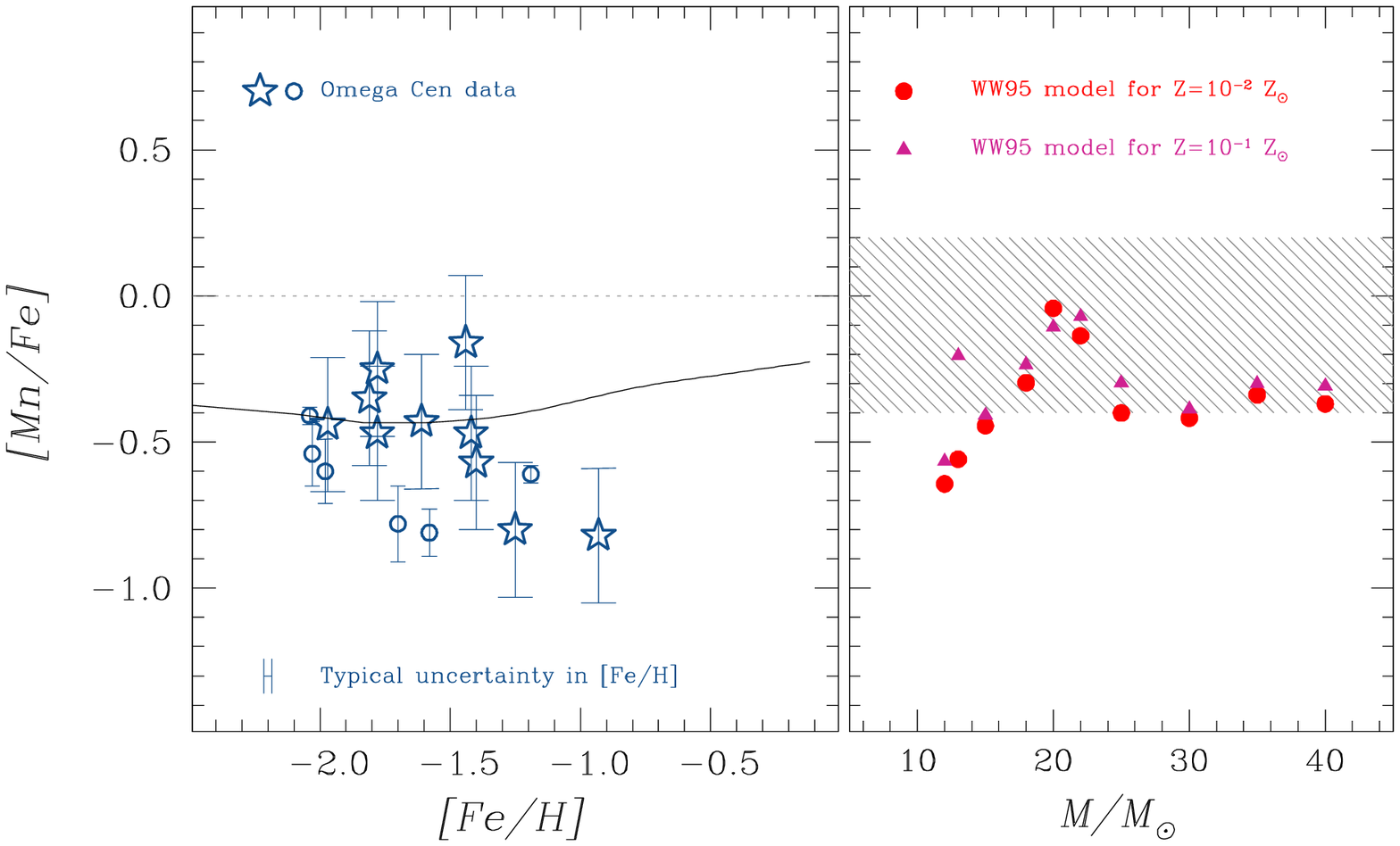,width=\textwidth}
     \caption{ Left panel: [Mn/Fe] versus [Fe/H] in $\omega$\,Cen as measured 
       in giants (stars: Cunha et al. 2010) and subgiants (empty circles: 
       Pancino et al. 2010) and predicted by our fiducial chemical evolution 
       model (solid line). Right panel: symbols: chemical composition of pure 
       SNII ejecta from individual massive stars from Woosley \& Weaver (1995), 
       for metallicities typical of the cluster's stars (filled circles: 
       $Z/Z_{\odot} = 10^{-2}$; filled triangles: $Z/Z_{\odot} = 10^{-1}$); 
       shaded area: [Mn/Fe] ratio in the ejecta of SNeIa for metallicities 
       typical of the cluster, according to Cescutti et al. (2008) recipe. The 
       masses of the binary systems leading to SNIa explosions go from 3 to 16 
       M$_{\odot}$.
     }
     \label{fig:disp}
   \end{figure*}
%
%%%%%%%%%%%%%%%%%%%%%%%%%%%%%%%%%%%%%%%%%%%%%%%%

   \subsection{The classical picture}

   The dashed curves in Fig.~\ref{fig:lte} show the behaviour of [Mn/Fe] versus 
   [Fe/H] predicted for the solar neighbourhood (left panel) and for 
   $\omega$\,Cen (right panel) with our models using the metallicity-dependent 
   yields of Woosley \& Weaver (1995) for SNeII and model W7 of Iwamoto et al. 
   (1999) for SNeIa at all metallicities. The theoretical predictions, in this 
   and all the following figures, are normalized by the solar values of 
   Grevesse \& Sauval (1998). Here, the results of the models are compared to 
   LTE data.

   It can be immediately seen that, as soon as SNeIa start to sensibly 
   contribute to the chemical enrichment of the ISM\footnote{According to the 
     time-delay model, this happens at different times (metallicities) in the 
     two systems, namely 0.5 Gyr from the beginning of the star formation 
     ([Fe/H]~$\sim -$1) in the solar neighbourhood and 0.1 Gyr from the 
     beginning of the star formation ([Fe/H]~$\sim -$2) in $\omega$\,Cen.}, the 
   theoretical curves begin to diverge from the observed [Mn/Fe] versus [Fe/H] 
   relation. This can be interpreted as an indication that the Mn yields from 
   SNeIa are overestimated (see also Cescutti et al. 2008). Indeed, by 
   introducing a metallicity dependence in the Mn yield from SNeIa (either by 
   interpolating between models W70 and W7 of Iwamoto et al. 1999 or by 
   adopting a Mn yield decreasing with decreasing metallicity as in Cescutti et 
   al. 2008 -- see Eq.~\ref{eq:MnZ}), we get a better fit to the solar 
   neighbourhood data (left panel, dotted and solid lines, respectively). 
   However, the low [Mn/Fe] ratios observed in the most metal-rich stars of 
   $\omega$\,Cen cannot be recovered by the model, which always predicts a 
   [Mn/Fe] ratio increasing with time (metallicity) in $\omega$\,Cen (right 
   panel, dotted and solid lines).

   \subsection{Turning to non-LTE abundances}

   The analysis of the evolution of several element-to-iron abundance ratios in 
   the Milky Way led Timmes, Woosley \& Weaver (1995) to favour the use of 
   \emph{half the nominal values} of the Fe yields given by Woosley \& Weaver 
   (1995) in chemical evolution models. Therefore, we have recomputed our set 
   of chemical evolution models by reducing the Woosley \& Weaver (1995) 
   original Fe yields by a factor of 2. The results are shown in 
   Fig.~\ref{fig:nlte}, for either metal-independent (dashed curves) or 
   metal-dependent (dotted and solid curves) Mn yields from SNeIa. The 
   theoretical predictions are compared to non-LTE abundance data. The non-LTE 
   corrected Mn abundances of metal-poor stars are higher than the 
   corresponding LTE values. Therefore, a very good agreement is obtained 
   between the observed and predicted [Mn/Fe] versus [Fe/H] trends for the 
   Galactic halo when assuming half the nominal values of the Fe yields by 
   Woosley \& Weaver (1995) and the non-LTE corrected Mn abundances for halo 
   stars.
   For [Fe/H]~$> -$1.0, a rise to solar ratios shallower than 
   predicted should be probably preferred, though more data are needed to 
   characterize the run of [Mn/Fe] with [Fe/H] at disc metallicities. The 
   adoption of a metallicity-dependent yield of Mn from SNeIa produces a 
   constant rather than increasing [Mn/Fe] ratio in $\omega$\,Cen only when the 
   empirical law by Cescutti et al. (2008) is adopted (solid line). 
   Interpolating linearly between models W70 and W7 of Iwamoto et al. (1999), 
   instead, still results in a [Mn/Fe] ratio increasing with [Fe/H] (dotted 
   line). Even a constant [Mn/Fe] ratio, however, does not suffice to explain 
   the observations. Current measurements, in fact, seem to point to an abrupt 
   fall of [Mn/Fe] at [Fe/H]~$\sim -$1.2 in this cluster (Fig.~\ref{fig:nlte}, 
   right panel), though one must be aware that this trend is dictated by just 
   two stars. A decreasing [Mn/Fe] ratio in $\omega$\,Cen may be obtained by 
   assuming that SNeIa do not produce any Mn; however, in that case it is 
   impossible to reproduce the Milky Way data (cfr. models labeled Nuc~0, 
   dot-dashed lines in Fig.~\ref{fig:nlte}). This could imply that SNeIa in the 
   Milky Way and in $\omega$\,Cen have different progenitors, but the paucity 
   of Mn data for $\omega$\,Cen prevents us from drawing any firm conclusion.

   \subsection{Invoking cooling flows and/or field star capture}

   The extremely low Mn abundance measured for a few stars in $\omega$\,Cen is 
   not the only chemical peculiarity of this cluster. Some of its stars are, in 
   fact, enormously enriched in helium and $s$-process elements, while 
   characteristic anticorrelations exist among the abundances of particular 
   elements, similarly to what is found in other globular clusters, but at 
   variance with the Milky Way field at the same metallicities.

   In order to explain the peculiar patterns observed in the chemical 
   abundances of a significant fraction of Galactic globular cluster stars, 
   D'Ercole et al. (2008) have suggested that the ejecta of first-generation 
   asymptotic giant branch (AGB) stars collect in a cooling flow into the 
   cluster core, where they form a subsystem of chemically anomalous 
   second-generation stars. Though attractive, such a scenario cannot provide 
   a justification for the low Mn abundances ([Mn/Fe]~$\sim -$0.8) observed in 
   metal-rich stars of $\omega$\,Cen by Cunha et al. (2010). Indeed, 
   first-generation AGB stars in $\omega$\,Cen would display a value of [Mn/Fe] 
   close to $-$0.4 (or $-$0.2, depending on whether the SNII Fe yields are 
   taken at face value or reduced by a factor of 2) in their ejecta, dictated 
   by SNII nucleosynthesis at low metallicities. This value is well above that 
   suggested by the observations of relatively metal-rich stars in 
   $\omega$\,Cen. The same considerations apply to the competitive scenario 
   proposed by Decressin et al. (2007), where the chemical peculiarities of 
   second-generation stars are driven by the slow winds of rotating massive 
   stars.

   Alternatively, one might argue that chemically peculiar stars did not 
   originate in the cluster itself, but were accreted from the surroundings 
   (see Fellhauer, Kroupa \& Evans 2006, for a possible scenario of field stars 
   trapping by the newborn $\omega$\,Cen). In this case, they would be no 
   longer representative of the self-enrichment history of the cluster. 
   However, Mn abundances as low as [Mn/Fe] $\simeq -$0.8, as  measured in 
   $\omega$\,Cen, have never been detected elsewhere, which makes the field 
   star capture hypothesis highly unreliable. In Fig.~\ref{fig:all}, we compare 
   the [Mn/Fe] versus [Fe/H] trends observed in different stellar systems -- 
   the solar neighbourhood, several Galactic globular clusters and a few dSphs 
   -- to the [Mn/Fe] versus [Fe/H] relation of $\omega$\,Cen stars (notice that 
   part of the large scatter in the $\omega$\,Cen data has been artificially 
   introduced by plotting together measurements based on different Mn lines; 
   see discussion in Sect.~2 and Pancino et al. 2011). It is clearly seen that, 
   below [Fe/H]~= $-$0.8, all stars, independently of the system they belong 
   to, share a common value of [Mn/Fe] of roughly $-$0.4. This is not true only 
   for a few stars in $\omega$\,Cen with [Mn/Fe]~$\sim -$0.8.

   \subsection{Inhomogeneous chemical evolution?}

   At this point, it remains to be assessed whether we can explain the 
   observations by relaxing some of the simplifying hypotheses of our model. 
   
   The solid curve in Fig.~\ref{fig:disp}, left-hand panel, tracks down the 
   predicted behaviour of [Mn/Fe] as a function of [Fe/H] in the ISM of 
   $\omega$\,Cen, according to our fiducial model for this cluster. In the 
   framework of homogeneous chemical evolution, the chemical composition of the 
   newborn stars is exactly that predicted for the ISM at the time of their 
   formation. Any spread in the data is, therefore, left unexplained. 

   For [Fe/H]~$< -$1.4, the majority of the data points is consistent with the 
   predictions of our homogeneous chemical evolution model, within the quoted 
   uncertainties. For [Fe/H]~$> -$1.4, there are only 3 stars with Mn 
   determinations in $\omega$\,Cen, and all of them lie below the theoretical 
   curve. While this may be a hint for a decrease of [Mn/Fe] with [Fe/H] in 
   $\omega$\,Cen, it was pointed out to us that one should also consider the 
   possibility that the stars form from a medium that is not well-mixed and 
   thus bears the signature of chemical enrichment from a few SNe only.

   In Fig.~\ref{fig:disp}, right-hand panel, we show the composition of pure 
   SNII ejecta as a function of the initial stellar mass, for a grid of stellar 
   models from 12 to 40~M$_{\odot}$ (filled circles and triangles). We take the 
   yields from Woosley \& Weaver (1995), for two values of the metallicity 
   typical of $\omega$\,Cen's stars. If the star formation in $\omega$\,Cen's 
   progenitor is initiated by SNe and the stars form from a mixture of 
   `snowplowed' interstellar medium and individual supernova ejecta, in a 
   scenario resembling that suggested for the Galactic halo by Ishimaru \& 
   Wanajo (1999; see also Tsujimoto et al. 1999, Argast et al. 2000 and 
   Cescutti 2008, for inhomogeneous chemical evolution modeling of the halo), a 
   broad distribution of [Mn/Fe] in the newborn stars is expected, because of 
   the broad distribution of [Mn/Fe] ratios in the ejecta of individual 
   core-collapse SNe. However, the lowest [Mn/Fe] ratios observed in 
   $\omega$\,Cen are left unexplained. Also shown in Fig.~\ref{fig:disp} are 
   the [Mn/Fe] ratios in the ejecta of SNeIa that explode in $\omega$\,Cen's 
   progenitor (shaded area; the Mn yield from SNeIa is computed according to 
   Cescutti et al.'s 2008 recipe). Once again, it is seen that the stars with 
   the lowest [Mn/Fe] ratios in $\omega$\,Cen can not be explained as forming 
   from pure SN ejecta.

   Though relaxing the hypothesis of homogeneous chemical evolution seems a 
   promising way to obtain for some stars theoretical Mn abundances lower than 
   predicted for the gas, the model predictions can hardly be brought into 
   agreement with the observations of the most metal-rich stars. We also want 
   to emphasize the following: Inhomogeneous chemical evolution models for the 
   Galactic halo always predict that the spread in the data is reduced in the 
   course of the evolution of the system (see the references above), eventually 
   leading to convergence with the predictions from homogeneous models. 
   However, in $\omega$\,Cen we would see exactly the opposite, i.e. an 
   increase in the dispersion with time, \emph{unless the data for the most 
     metal-rich stars are tracing a genuine decrease} of the [Mn/Fe] ratio in 
   the cluster. This is an intriguing aspect, that should be further 
   investigated in more populous samples of high-metallicity $\omega$\,Cen 
   stars.

   \section{Conclusions}

   In this paper we present the theoretical [Mn/Fe] versus [Fe/H] relationships 
   predicted with our self-consistent chemical evolution model for 
   $\omega$\,Cen, by using different prescriptions on Mn and Fe synthesis in 
   stars. In particular, as for Mn production from SNeIa, we adopt either a 
   metal-independent yield, by assuming Iwamoto et al.'s (1999) W7 model at all 
   metallicities, or a metal-dependent one, by interpolating between models W70 
   and W7 of Iwamoto et al. (1999) or by adopting the empirical law of Cescutti 
   et al. (2008). We also run models without Mn production from SNeIa. The 
   theoretical relations are then compared to LTE and non-LTE data on Mn 
   abundances in $\omega$\,Cen giants.

   The adopted chemical evolution model reproduces all the major chemical 
   properties of $\omega$\,Cen -- its metallicity distribution function, 
   age-metallicity relation, average trends of several $\alpha$-element-to-iron 
   abundance ratios as functions of [Fe/H] (Romano et al. 2007). It also 
   accounts for the presence of extreme He-rich stars (in the right percentage) 
   and for the existence of a Na-O anticorrelation in the cluster (Romano et 
   al. 2010a). However, the trend of decreasing [Mn/Fe] with increasing [Fe/H], 
   displayed by both LTE and non-LTE data, is not reproduced by the model, 
   which predicts instead a [Mn/Fe] ratio either increasing or constant in 
   time, depending on the choice of stellar yields. The adoption of a 
   metallicity-dependent, rather than metal-independent, Mn yield from SNeIa 
   leads to an almost flat [Mn/Fe] versus [Fe/H] trend in $\omega$\,Cen only if 
   the empirical law of Cescutti et al. (2008) is adopted. By interpolating 
   between models W70 and W7 of Iwamoto et al. (1999), i.e. between models 
   computed for metal-free and solar-metallicity SNeIa, respectively, we still 
   produce a [Mn/Fe] ratio increasing with [Fe/H] in $\omega$\,Cen. On the 
   basis of the discussion in Sect.~4 of the present paper and on previous 
   results by Cescutti et al. (2008), we conclude that a flat trend has to be 
   preferred over an increasing one. 

   In our chemical evolution model for $\omega$\,Cen, the chemical properties 
   of the cluster are mainly driven by the action of strong differential 
   galactic winds, which deeply affect the evolution of its dSph precursor (see 
   Romano et al. 2007, and references therein, for details on the adopted 
   scenario). Allowing for cooling flows or field star capture would not help 
   to explain the presence of low-Mn stars in the cluster. Relaxing the 
   hypothesis of homogeneous chemical evolution, i.e. allowing the stars to 
   form from a mixture of ISM and individual SN ejecta, could eventually lead 
   (for some stars) to the prediction of Mn abundances lower than predicted for 
   the gas. However, this would hardly accommodate the Mn abundances of the two 
   most metal-rich stars in Cunha et al.'s (2010) sample in a scenario of flat 
   -- rather than decreasing -- Mn evolution with time.

   We suggest that more measurements of Mn in $\omega$\,Cen stars at high 
   metallicity are needed to finally set the issue of Mn evolution in 
   $\omega$\,Cen. Interestingly, [Mn/Fe] in Sagittarius stays almost flat 
   (Sbordone et al. 2007; see Fig.~\ref{fig:all}) and Sagittarius is likely to 
   be the local counterpart of the accretion episode which led to the formation 
   of $\omega$\,Cen some 10 Gyr ago (Bellazzini et al. 2008; Georgiev et al. 
   2009; Carretta et al. 2010). Hence, it would be not surprising if future 
   measurements of Mn in a much bigger sample of $\omega$\,Cen stars should 
   reveal a constant run of Mn with [Fe/H], rather than the decreasing trend 
   suggested by Cunha et al. (2010) on the basis of extant data. On the other 
   hand, would future measurements confirm a fall of the [Mn/Fe] ratio towards 
   higher metallicities, a revision of current scenarios of the formation of 
   the cluster may be needed, since none of them is able to explain a 
   decreasing trend of Mn in $\omega$\,Cen. We have shown that, if Mn 
   production from SNeIa is totally suppressed in $\omega$\,Cen, a [Mn/Fe] 
   ratio decreasing with time (metallicity) can be found in the cluster. While 
   it is pointless to speculate on this theoretical result until more Mn data 
   for $\omega$\,Cen stars become available, we note that Jonhson \& 
   Pilachowski (2010) observe consistently elevated [$\alpha$/Fe] ratios for 
   nearly all stars in the cluster (their sample totals 855 giants) and 
   interpret this as evidence against a significant contribution to 
   $\omega$\,Cen's chemical enrichment from SNeIa.

   \section*{Acknowledgments}
   We thank E.~Carretta for providing unpublished data on Mn abundances in 
   Galactic globular clusters. One of us (FM) thanks Prof. F.-K.~Thielemann for 
   illuminating discussions on nucleosynthesis in SNeIa. DR thanks V.~Hill and 
   E.~Tolstoy for comments on preliminary results from this work presented at 
   the International Space Science Institute (ISSI) in Bern (CH) in the context 
   of Team meetings. The hospitality and financial support of ISSI are also 
   gratefully acknowledged. We acknowledge support from Italian MIUR (Ministero 
   dell'Istruzione, dell'Universit\`a e della Ricerca) through grant PRIN~2007, 
   prot.~no.~2007JJC53X\_001 and INAF through grant PRIN~2009 ``Formation and 
   Early Evolution of Massive Star Clusters''.

\bsp

\label{lastpage}

\end{document}